\documentclass[%
 aip,
 amsmath,amssymb,
 reprint,%
]{revtex4-2}

\usepackage{graphicx}
\usepackage{dcolumn}
\usepackage{bm}

\usepackage[utf8]{inputenc}
\usepackage[T1]{fontenc}
\usepackage{mathptmx}
\usepackage{color}
\usepackage{ulem}

\usepackage{placeins}

\begin{document}

\preprint{AIP/123-QED}

\title{Protocol for temperature sensing using a three-level transmon circuit}

\author{Aidar Sultanov}
\affiliation{QTF  Centre  of  Excellence,  Department  of  Applied  Physics, School  of  Science,  Aalto  University,  FI-00076  Aalto,  Finland}%

\author{Marko Kuzmanovi\'{c}}%
\affiliation{QTF  Centre  of  Excellence,  Department  of  Applied  Physics, School  of  Science,  Aalto  University,  FI-00076  Aalto,  Finland}%

\author{Andrey V. Lebedev}
\affiliation{Dukhov Research Institute of Automatics (VNIIA), Moscow 127055, Russia}
\affiliation{Moscow  Institute  of  Physics  and  Technology,  141700, Institutskii  Per.  9,  Dolgoprudny,  Moscow  Distr.,  Russia}

\author{Gheorghe Sorin Paraoanu}
\email{sorin.paraoanu@aalto.fi}
\affiliation{QTF  Centre  of  Excellence,  Department  of  Applied  Physics, School  of  Science,  Aalto  University,  FI-00076  Aalto,  Finland}%

\date{\today}

\begin{abstract}
\section*{Abstract}
We present a method for {\it in situ} temperature measurement  of superconducting quantum circuits, \textcolor{black}{by using the first three levels of a transmon device to which we apply a sequence of $\pi$ gates.}
Our approach employs projective dispersive readout and utilizes the basic properties of the density matrix associated with thermal states. \textcolor{black}{This method works with an averaging readout scheme and does not require a single-shot readout setup.} We validate this protocol by performing thermometry in the range of 50 mK - 200 mK, corresponding to a range of residual populations $1\%-20 \%$  for the first excited state and $0.02\%-3 \%$ for the second excited state. 
\end{abstract}

\maketitle


Superconducting qubits are one of the most promising candidates as the basic element of future quantum computers. The progress of the last decade \textcolor{black}{has} resulted in a significant increase of their coherence times to tens of microseconds \cite{Rigetti2012, Oliver2013, Paik2011}, \textcolor{black}{in a reduction of  errors caused by interaction with the environment through the implementation of reset protocols  \cite{Geerlings2013, Valenzuela2006, Tuorila2017, Riste2012, Egger2018, Magnard2018} and error-correction protocols 
\cite{ Campagne2020,Reed2012}}, and in an enhancement in readout fidelity up to $99.6 \%$ \cite{Vijay2011,deLange2014,Lin2013,Abdo2014}. However, the exact mechanisms that limits further improvements in superconducting qubit systems are still not fully understood; one possibility is the spurious excitations caused by microwave noise, infrared radiation from hotter stages of the dilution refrigerators or poisoning by quasiparticles \cite{Corcolesa2011,Wenner2013,Serniak2018,Barends2011}. To mitigate these effects, a range of experimental techniques have been deployed -- the use of cryogenic filters and attenuators, infrared absorbers, radiation and magnetic shielding of samples, with the goal of reducing the temperature of the electromagnetic environment and the quasiparticle population.
Here we introduce a protocol for evaluating the effective temperature of a superconducting qubit. Our method can be readily used as a diagnostic tool for qubit thermalization and line integrity in quantum computing applications. An important application is quantum thermodynamics \cite{Pekola2015,Silveri2017,Marco2021}, where controlling the effective temperature of the circuit can be used to drive quantum engines.

The state of the electromagnetic environment of the qubit is described by an effective temperature, which characterizes the thermal equilibrium between the qubit and the environment and thus defines residual populations of former.
There are several ways to estimate this effective temperature from the residual populations of qubit's states, assuming a Maxwell-Boltzmann distribution. A straightforward method is to use a  single-shot readout. In this case the residual probabilities can be directly calculated from measurement statistics, \textcolor{black}{provided that the states can be discriminated with sufficiently good precision}.  However, the implementation of a single-shot readout scheme requires a good quantum limited parametric amplifier and additional components \cite{Johnson2012,Riste2012_2,Krantz2016}. An alternative  approach, which does not use single-shot readout, is based on the measurement of correlations between responses corresponding to the ground and excited states \cite{Kulikov2020}. Another technique uses a three level system, where the Rabi oscillation amplitude between the first and the second excited state depends on the residual population of the first excited state \cite{Geerlings2013,Jin2015}. However, this method is highly sensitive to the readout signal parameters. 
Finally, a thermometry technique for propagating waves in open-waveguides \cite{Simone2020} can be used to characterize the temperature of the electromagnetic field, but this method requires a dedicated sample design.

Here we propose an {\it in situ} method for measuring the effective temperature, which utilizes only $\pi$ pulses and requires measuring only the average responses in the dispersive readout limit. Therefore this method could be implemented without a specialized setup  or  \textcolor{black}{sophisticated} measurement techniques. 
In addition, determining the temperature does not rely on qubit state tomography: In our protocol, we measure the cavity responses after applying six different drive sequences that swap the populations of the three-level system, in our case defined by the first levels in a transmon device. A simple linear relationship is found between some of these responses, and the coefficient of proportionality is determined only by the thermal level occupations.  Therefore, as the method does not rely on \textcolor{black}{full state} tomography or on the knowledge of the pure state responses, it is more resilient to noise and drifts which are commonly present in superconducting artificial atom experiments. Moreover, since only $\pi$ pulses are utilized, the proposed method is robust against dephasing and, if the pulses are much shorter with respect to the relaxation time, also against decay. 


Consider a three-level system in thermal equilibrium with its environment at a temperature $T$. The density matrix reads
\begin{equation}\label{state}
\hat{\rho}=p_g\left|\left.g\right\rangle\right.\left\langle\left.g\right|+p_e\left|\left.e\right\rangle\right.\left\langle\left.e\right|\right.+p_f\left|\left.f\right\rangle\right.\left\langle\left.f\right|\right.\right.,
\end{equation}
where  $\left|\left.g\right\rangle\right.$, $\left|\left.e\right\rangle\right.$, $\left|\left.f\right\rangle\right.$ are respectively the ground, the first excited and the second excited state, with corresponding populations $p_g$, $p_e$, and $p_f$. Thermal equilibrium means that $\hat{\rho}$ is diagonal and the residual populations are defined by the Maxwell-Boltzmann distribution: 
\begin{equation}\label{distrib}
p_g=\frac{1}{Z}e^{-\frac{E_g}{k_{\rm B}T}},~~
p_e=\frac{1}{Z}e^{-\frac{E_e}{k_{\rm B}T}},~~
p_f=\frac{1}{Z}e^{-\frac{E_f}{k_{\rm B}T}}
\end{equation}
where $k_{\rm B}$ is the Boltzmann constant, $T$ is the effective temperature, $E_i$ with $i\in \{g,e,f\}$ are the energies of the corresponding states and \textcolor{black}{ $Z=\sum_{i}\exp\left[-E_i/k_BT\right]$} is the canonical partition function.

For a transmon device, the readout of these three levels is implemented through the projective  measurement operators \cite{Biancetti2010}
\begin{subequations}
 \begin{align}
  \widehat{M}_I={\varphi}_g^I\left|\left.g\right\rangle\right.\left\langle\left.g\right|\right.\ +{\varphi}_e^I\left|\left.e\right\rangle\right.\left\langle\left.e\right|\right.+{\varphi}_f^I\left|\left.f\right\rangle\right.\left\langle\left.f\right|\right.,  \label{MeasopI}  \\
  \widehat{M}_Q={\varphi}_g^Q\left|\left.g\right\rangle\right.\left\langle\left.g\right|\right.+{\varphi}_e^Q\left|\left.e\right\rangle\right.\left\langle\left.e\right|\right.+{\varphi}_f^Q\left|\left.f\right\rangle\right.\left\langle\left.f\right|\right. ,\label{MeasopQ}
 \end{align}
\end{subequations}
where $\widehat{M}_I$, $\widehat{M}_Q$ are measurement operators, corresponding to the $I$ and $Q$ quadratures of the measured signal: These quadratures are denoted by ${\varphi}_i^{I(Q)}$ for the corresponding states.  More precisely, in this formalism, ${\varphi}_i^{I(Q)}$ is the $I(Q)$  quadrature of the measured signal if the device is prepared in the state $|i\rangle$. Note that ${\varphi}_i^{I(Q)}$ are time-dependent functions, which makes the operators $\widehat{M}_{I(Q)}$ also time-dependent.

The averaged measurement result of an arbitrary state is defined as follows: 

\begin{subequations}
 \begin{align*}
\left\langle\left.I\right\rangle\right.=  Tr\left(\hat{\rho}\widehat{M}_I\right) , \\
\left\langle\left.Q\right\rangle\right.=Tr\left(\hat{\rho}\widehat{M}_Q\right).
 \end{align*}
\end{subequations}

For a thermal state $\hat{\rho}$, the measurement outcome becomes:
\begin{subequations}\label{measures}
 \begin{align}
\left\langle\left.I\right\rangle\right.=  {p_g{\varphi}}_g^I +p_e{\varphi}_e^I+p_f{\varphi}_f^I\label{OutcomeI} , \\
\left\langle\left.Q\right\rangle\right.={p_g{\varphi}}_g^Q +p_e{\varphi}_e^Q  +p_f{\varphi}_f^Q\label{OutcomeQ} .
 \end{align}
\end{subequations}

If the pure state responses $\varphi^{I/Q}_{g,e,f}$ were known, one could in principle extract the thermal populations by linear regression. However, in the averaged readout scheme only the ensemble average is accessible. 

To overcome this difficulty, we propose to measure the responses after applying  certain pulse sequences that swap the populations of three level systems in the density matrix Eq. (\ref{state}). As we will see, the protocol allows us to eliminate completely the unknown responses $\varphi_{i}^{I(Q)}$.
Let us denote the pulse swapping the ground and first excited state populations as $\pi_{ge}$ and that swapping the first and second excited states as $\pi_{ef}$. 
All used sequenced are summarized in Table \ref{Sequences}. For example, when a single $\pi_{ge}$ pulse is applied, one gets the state $\hat{\rho}=p_e\left|\left.g\right\rangle\right.\left\langle\left.g\right|+p_g\left|\left.e\right\rangle\right.\left\langle\left.e\right|\right.+p_f\left|\left.f\right\rangle\right.\left\langle\left.f\right|\right.\right.$, and according to Eq. (\ref{measures}), we get the output of the $I$-quadrature as ${p_e{\varphi}}_g^I\left(t\right)\ +p_g{\varphi}_e^I\left(t\right)+p_f{\varphi}_f^I\left(t\right)$. 
 \textcolor{black}{ Here, we note that, in order  to implement the protocol and the proposed sequences of gates, the second excited state should be accessible by dispersive readout.} 

\begin{table}
\centering

\begin{tabular}{|m{1.8cm}|  p{5.7cm}|p{1cm}|}
\hline
 \textbf{Sequence} &  \textbf{Outcome}  & \textbf{Label}  \\
\hline
no pulses &$ {p_g{\varphi}}_g^{I(Q)} +p_e{\varphi}_e^{I(Q)} +p_f{\varphi}_f^{I(Q)}$ & $x_0^{I(Q)}$ \\[2ex]
\hline
$\pi_{ge}$  &$ {p_e{\varphi}}_g^{I(Q)} +p_g{\varphi}_e^{I(Q)}+p_f{\varphi}_f^{I(Q)}$ & $x_1^{I(Q)}$ \\ [2ex]
\hline
$\pi_{ge}$ $\pi_{ef}$ &$ {p_e{\varphi}}_g^{I(Q)} +p_f{\varphi}_e^{I(Q)}+p_g{\varphi}_f^{I(Q)}$ & $x_2^{I(Q)}$ \\ [2ex]
\hline
$\pi_{ef}$ &$ {p_g{\varphi}}_g^{I(Q)} +p_f{\varphi}_e^{I(Q)}+p_e{\varphi}_f^{I(Q)}$ & $y_0^{I(Q)}$ \\ [2ex]
\hline
$\pi_{ef}$$\pi_{ge}$ &$ {p_f{\varphi}}_g^{I(Q)} +p_g{\varphi}_e^{I(Q)}+p_e{\varphi}_f^{I(Q)}$ & $y_1^{I(Q)}$ \\ [2ex]
\hline
$\pi_{ef}$$\pi_{ge}$ $\pi_{ef}$  &$ {p_f{\varphi}}_g^{I(Q)} +p_e{\varphi}_e^{I(Q)}+p_g{\varphi}_f^{I(Q)}$ & $y_2^{I(Q)}$ \\ [2ex]
\hline
\end{tabular}
\caption{Sequences of operations used for the temperature measurement protocol.}
\label{Sequences}
\end{table}

 \textcolor{black}{In general, the responses $\varphi(t)_{g,e,f}=\varphi(t)_{g,e,f}^I+i\varphi(t)_{g,e,f}^Q$ can be understood as vectors in an infinite-dimensional (with respect to time) vector space over a complex $I+iQ$ field. For the sake of simplicity, from now on we present all the expressions for the $I$ and $Q$ components separately. From this point of view, the differences of some of these responses can be classified according to the collinearity criterion.} For example, the difference of $x_0^{I\left(Q\right)}-x_1^{I\left(Q\right)}=(p_g-p_e) ({\varphi}_g^{I(Q)}-{\varphi}_e^{I(Q)})$   and 
$y_0^{I\left(Q\right)}-y_1^{I\left(Q\right)}=(p_g-p_f) ({\varphi}_g^{I(Q)}-{\varphi}_e^{I(Q)})$ can be seen as two collinear vectors in the space spanned by $\varphi_{g,e,f}^{I,Q}$ and which lie along the direction $\varphi_{ge}={\varphi}_g^{I(Q)}-{\varphi}_e^{I(Q)}$.
Therefore ${x_0^{I\left(Q\right)}-x_1^{I\left(Q\right)}}=({y_0^{I\left(Q\right)}-y_1^{I\left(Q\right)}})\frac{p_g-p_e}{p_g-p_f}$, and it is possible to determine the coefficient of proportionality $A^{I\left(Q\right)}=\frac{p_g-p_e}{p_g-p_f}$ along the direction $\varphi_{ge}$, without knowledge of the pure state responses. Similarly, $A$ is also the slope between either ${y_0^{I\left(Q\right)}-x_2^{I\left(Q\right)}}$  and ${x_0^{I\left(Q\right)}-y_2^{I\left(Q\right)}}$, along the direction $\varphi_{gf}$; or the slope between ${y_1^{I\left(Q\right)}-y_2^{I\left(Q\right)}}$ and ${x_1^{I\left(Q\right)}-x_2^{I\left(Q\right)}}$ along the direction $\varphi_{ef}$.

Overall,we have identified the following pairs of differences: 
\begin{equation}\label{slopes}
\left\{\begin{matrix}
A=\frac{x_0^{I\left(Q\right)}-x_1^{I\left(Q\right)}}{y_0^{I\left(Q\right)}-y_1^{I\left(Q\right)}}=\frac{y_0^{I\left(Q\right)}-x_2^{I\left(Q\right)}}{x_0^{I\left(Q\right)}-y_2^{I\left(Q\right)}}=\frac{y_1^{I\left(Q\right)}-y_2^{I\left(Q\right)}}{x_1^{I\left(Q\right)}-x_2^{I\left(Q\right)}}=\frac{p_g-p_e}{p_g-p_f},\\
B=\frac{x_1^{I\left(Q\right)}-y_1^{I\left(Q\right)}}{y_0^{I\left(Q\right)}-x_2^{I\left(Q\right)}}=\frac{x_2^{I\left(Q\right)}-y_2^{I\left(Q\right)}}{x_0^{I\left(Q\right)}-x_1^{I\left(Q\right)}}=\frac{x_0^{I\left(Q\right)}-y_0^{I\left(Q\right)}}{y_1^{I\left(Q\right)}-y_2^{I\left(Q\right)}}=\frac{p_e-p_f}{p_g-p_e} ,\\\end{matrix}\right.
\end{equation}
where column-wise ratios of responses are given  along the directions of $\varphi_{ge}$,  $\varphi_{gf}$ and $\varphi_{ef}$, correspondingly. 

The coefficients of proportionality $A$ and $B$ are uniquely determined by the temperature $T$ and the transition frequencies $\hbar\omega_{ge}=E_g-E_e$ and $\hbar\omega_{gf}=E_g-E_f$:

\begin{equation}\label{slopes_fun_temp}
\left\{\begin{matrix} A=\frac{1-\exp\left({\hbar\omega_{ge}}/{k_{\rm B}T}\right)}{1-\exp\left({\hbar\omega_{gf}}/{k_{\rm B}T}\right)},\\
B=\frac{\exp\left({\hbar\omega_{ge}}/{k_{\rm B}T}\right)-\exp\left({\hbar\omega_{gf}}/{k_{\rm B}T}\right)}{1-\exp\left({\hbar\omega_{ge}}/{k_{\rm B}T}\right)}.\end{matrix}\right.
\end{equation}

From Eq. (\ref{slopes_fun_temp}) it is possible to determine the temperature $T$, assuming that the transition frequencies are known.

Note that in principle one could introduce also the coefficient
\begin{equation}
C=\frac{x_1^{I\left(Q\right)}-y_1^{I\left(Q\right)}}{x_0^{I\left(Q\right)}-y_2^{I\left(Q\right)}}=\frac{x_2^{I\left(Q\right)}-y_2^{I\left(Q\right)}}{y_0^{I\left(Q\right)}-y_1^{I\left(Q\right)}}=
\frac{x_0^{I\left(Q\right)}-y_0^{I\left(Q\right)}}{x_1^{I\left(Q\right)}-x_2^{I\left(Q\right)}}.
\end{equation}
However, it is easily verified that
\begin{equation}
C=\frac{p_e-p_f}{p_g-p_f}=\frac{\exp\left({\hbar\omega_{ge}}/{k_{\rm B}T}\right)-\exp\left({\hbar\omega_{gf}}/{kT}\right)}{1-\exp\left({\hbar\omega_{gf}}/{k_{\rm B}T}\right)},
\end{equation}
therefore $C=A*B$, and therefore it does not provide an independent measure of the temperature, but can be used to estimate the accuracy of the protocol, which is discussed further in the Supplementary Material.
Finally, we note that the protocol uses only $\pi$ pulses, therefore it should be insensitive to qubit dephasing and to a large extent also to qubit relaxation, since the duration of the $\pi$ pulse is typically much smaller than the $T_1$ time.


 \textcolor{black}{Based on this protocol,} we implement two experiments using a standard low-temperature cryogenic setup with microwave wiring for input lines and with a heterodyne readout. 
The samples consist of a transmon device coupled to a microwave coplanar resonator, which is used for dispersive readout. The energy levels of the transmon  can be flux-tuned through a filtered DC flux bias line. The sample is thermally anchored to the mixing chamber of a dilution refrigerator and isolated from the output line by two circulators working in the $4-8$ GHz range. The readout pulse duration is 2 $\mu$s and the transmitted signal is amplified by 30 dB with a LNF amplifier at the 4.2 K  stage and by room-temperature amplifiers  by 60 dB. After the demodulation \textcolor{black}{to an IF frequency of 50 MHz,} the $IQ$ components are amplified by a low-bandwidth amplifier and digitized with 1 ns resolution with a data acquistion board (more information in the Supplementary Material). We calibrate the $\pi$ pulses for e-g and e-f transitions by standard Rabi experiments  and we use them to define the sequences of the population swapping operations described in Table \ref{Sequences}. 
The excitation pulses \textcolor{black}{are}  generated by an arbitrary waveform generator and mixed with a local oscillator (LO) frequency in an $IQ$ mixer. 

 \textcolor{black}{The pulses have Gaussian envelope and a duration between 56 ns and 120 ns, and the cross-excitation due to low transmon anharmonicity $f_{ge}-f_{ef} \approx 300$ MHz was negligible.}  We measure the response for each of the sequences, obtaining the IQ values of readout pulse. Each response is measured $60000$ times and averaged. The total attenuation is the same for both experiments and adds up to 75 dB  and 73 dB for the readout and drive lines correspondingly. 
In the main text the temperatures are determined using only the $I$ quadrature data, while in the Supplementary Material, for the error estimations  both the $I$ and $Q$ data are used.

The goal of the first experiment is to test the protocol. We use a sample with $E_{\rm c}/(2\pi )$ = 360 MHz and \textcolor{black}{$E_{\rm J}^{\rm max}/(2\pi )$ = 10.013 GHz},
and the readout resonator at $f_r=7.75$ GHz. \textcolor{black}{For the drive and readout lines we use an attenuation of 30 dB at the mixing chamber stage (MXC).} The effective temperature \textcolor{black}{is} measured as a function of flux bias \textcolor{black}{at fixed base MXC temperature.}

\begin{figure}[!ht]
\center
\includegraphics[width=\linewidth]{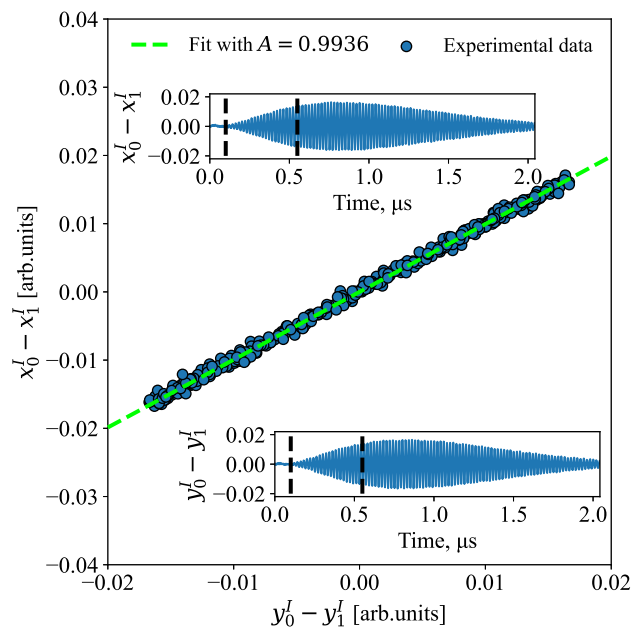}
\caption{\label{linearity_demo} Experimental verification of the linearity implied in Eq. (\ref{slopes}).  The bright green line is obtained by the linear regression algorithm resulting in a slope $A^{I}=0.9936$. The plots show the differences between the $I$-quadrature time-domain traces $x_0^I-x_1^I$ and $y_0^I - y_1^I$, with the range of data used for extracting the slope shown by blue dashed lines in the insets. \textcolor{black}{The raw data of readout signals are shown in the Supplementary Material.}}
\end{figure}

The measured coefficient $A^{I}$ is shown in Fig. \ref{linearity_demo}. In this figure, each point is obtained from a measurement of four readout traces $x_0^I, x_1^I,y_0^I$ and $y_1^I$ at a certain time. The differences $x_0^I-x_1^I$ and $y_0^I - y_1^I$ are shown in the inset, where the oscillations at the IF frequency clearly visible, with the time interval used for the linearity check delineated by dashed lines; further examples of measured responses could be found in the Supplementary Materials.
To extract the slopes we \textcolor{black}{implement} the Deming regression approach \cite{deming1943}, which considers noise in both X and Y axis. This approach relies on \textcolor{black}{the} assumption that errors in two-variable models are independent and follow a normal distribution law. We find that the extracted slope value $0.9936$ is very close to 1, as one could check from Eq. (\ref{slopes_fun_temp}), which implies that we expect a low temperature, since $\mathop {\lim }\limits_{T \to {0^ + }}A=
\mathop {\lim }\limits_{T \to {0^ + }} \frac{{1 - \exp \left( {{{\hbar {\omega _{ge}}} \mathord{\left/
 {\vphantom {{\hbar {\omega _{ge}}} {{k_B}T}}} \right.
 \kern-\nulldelimiterspace} {{k_B}T}}} \right)}}{{1 - \exp \left( {{{\hbar {\omega _{gf}}} \mathord{\left/
 {\vphantom {{\hbar {\omega _{gf}}} {{k_B}T}}} \right.
 \kern-\nulldelimiterspace} {{k_B}T}}} \right)}} = 1
$.

In Fig. \ref{Temperatures_lin} we show the extracted temperatures as a function of flux bias (qubit frequency). The results are in a good agreement with the base temperature of refrigerator and lay in the range of earlier reported temperatures, measured by other methods \cite{Johnson2012,Riste2012_2,Krantz2016,Geerlings2013,Jin2015,Kulikov2020}.
In thermal equilibrium with the environment, characterized by flat spectrum, the temperature should not depend on the transmon frequency \cite{Clerk2010}.
We see only a small variation of temperature, which proves that Sample 1 is generally well thermalized and the applied pulse sequences do not influence it significantly. The slight dependence of the effective temperature could be explained by the frequency dependent attenuation of \textcolor{black}{the} control lines. The spikes near $\omega_{ge}/2\pi=5.6-5.7$ GHz are most likely artifacts due to an imperfect calibration \textcolor{black}{of $\pi$ pulses}. 

The validity of our method \textcolor{black}{is} verified by a simulation of the system, where we model \textcolor{black}{the} Lindblad master equation with Boltzmann distribution for thermal photons; more details \textcolor{black}{are presented} in the Supplementary Material.  
\begin{figure}[!ht]
\includegraphics[width=\linewidth]{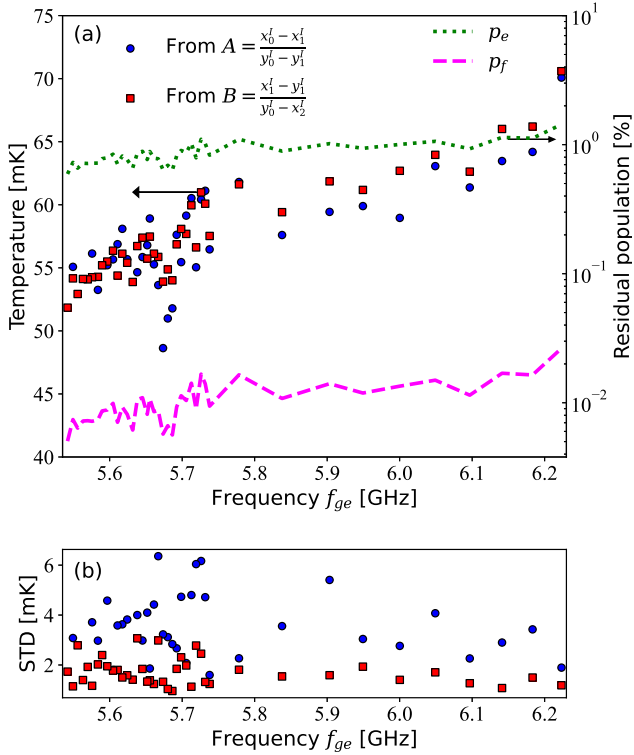}
\caption{\label{Temperatures_lin} \textcolor{black}{(a)} Effective temperature extracted from the responses $A$ (blue circles)  and $B$ (red squares), see Eq. (\ref{slopes}),  for Experiment 1 together with the residual populations of the state $|e\rangle$ (green dots) and $|f\rangle$ (magenta dashes). Note that the residual population on $|f\rangle$ remains below $0.03 \%$. \textcolor{black}{(b)} Standard deviation of the effective temperature obtained from 6 realizations of each measurement.}
\end{figure}


The goal in the second experiment is to demonstrate that the effective temperature can be controlled relatively independently from the temperature of the mixing chamber (MXC). The motivation comes from quantum thermodynamics, where  superconducting-circuit based Otto engines \cite{Karimi2016} and Stirling engines \cite{Sina2021} have been proposed theoretically. In these experiments it would be useful to have access to and set in a straightforward way the temperature of two reservoirs, the hot and the cold one. Here we show that by appropriate wiring we can have a relatively high temperature for the transmon, while at the same time maintaining the MXC as the cold bath. 

For these measurements we \textcolor{black}{have used} a sample with $E_{\rm c}/(2\pi )$ = 350 MHz and \textcolor{black}{$E_{\rm J}^{\rm max}/(2\pi )$  = 20.412 GHz}, while the resonator frequency is $f_r = 4.906$ GHz. To achieve a higher effective temperature,  
the previous 30 dB of attenuation in the input line at the mixing chamber stage \textcolor{black}{has been reduced by 15 dB, which results in a worse thermalization of the line and 
exposes the qubit to the thermal and non-equilibrium noise coming from the upper stages \cite{Krinner2019}. We observe an increase of the effective temperature of the transmon to about 160 mK.}

\begin{figure}[!ht]
\includegraphics[width=0.5\textwidth]{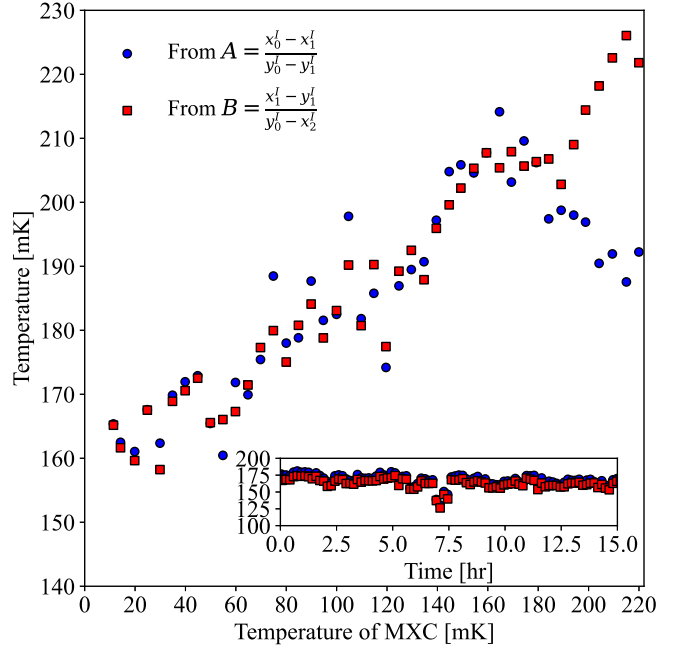}
\caption{\label{EffectiveT_vs_Tmxc} The effective temperature in Experiment 2 \textcolor{black}{as a} function of the base stage temperature (MXC). The inset shows the monitoring of effective temperature at a MXC temperature of 13 mK over 15 hours, showing the appearance of jumps.}
\end{figure}

Next, the effective temperature sensed by this sample is measured as a function of the base stage temperature; the results are shown in Fig. \ref{EffectiveT_vs_Tmxc}. We observe that the effective qubit temperature increases linearly with the MXC temperature. We have found that the slope in this linear dependence is approximately 1/3, therefore $T_{\rm eff} \approx T_{\rm MXC}/3 + 155$ mK. 

In the inset of Fig. \ref{EffectiveT_vs_Tmxc}, we present the results of monitoring the effective temperature at a fixed MXC temperature of 13 mK over 15 hours. The temperature is roughly constant, except for the observation of a switching event at 7.5 hours, most likely similar to the ones reported \textcolor{black}{before in the literature} \cite{Grunhaupt2018,Riste2013,Serniak2018,Weides2019,Bylander2019,Combes2021}.

\begin{figure}[!ht]
\includegraphics[width=0.5\textwidth]{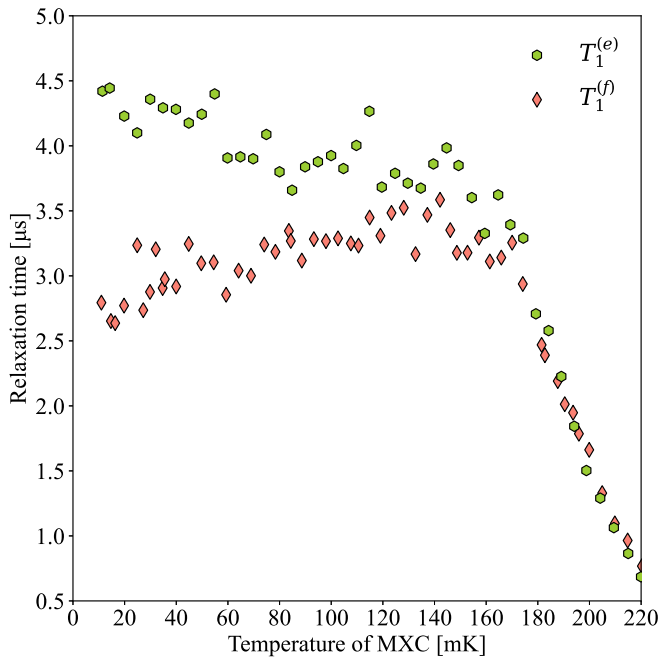}
\caption{\label{T1_vs_Temp} The relaxation times of the first and second excited states, $T_{1}^{(e)}$ (green hexagonal symbol) and $T_{1}^{(f)}$ (salmon diamond symbol) as a function of MXC temperature.}
\end{figure}

Above $T_{\rm MXC} \approx 170$ mK the effective temperatures estimated by $A$ and $B$ diverge. This can be understood as a consequence of the decreasing relaxation and coherence times at finite temperatures, and as a consequence the fidelities of drive and readout pulses \textcolor{black}{decrease}. To support this claim, we perform measurements of relaxation times. In Fig. \ref{T1_vs_Temp} the relaxation  times of the first and second excited states as a function of MXC temperatures are shown. Indeed, at MXC temperatures above  $170$ mK the $T_1$ times drop significantly, which roughly coincides with the start of divergence seen  in  Fig. \ref{EffectiveT_vs_Tmxc}. This behavior is well explained by models taking into account quasiparticle generation, see {\it e.g.}  \cite{Serniak2018,Catelani2011}, which predict a drop in $T_1$ at temperatures very close to what we see in Fig. \ref{T1_vs_Temp}. A slight increase in the relaxation time for the second excited state \textcolor{black}{has been observed in other experiments} \cite{Martinis2009}, and it \textcolor{black}{is} explained by non-equilibrium quasiparticles.

In summary, we \textcolor{black}{have proposed and demonstrated an} {\it in situ} method to extract the temperature by applying a sequence of $\pi$ gates to the first three levels of a transmon. The protocol is based on a standard setup, employing averaged readout for the transmon  states, does not require single-shot measurements, and it is robust against certain types of noise, such as relaxation and decoherence. The extracted temperatures are in the expected range, agreeing with previously reported effective temperatures of the same type of qubits. We have also shown that this allows for either the diagnosis of thermal radiation coming from the hotter stages of the fridge, or for the use of this radiation as a thermal reservoir for thermodynamic quantum engines.\\

\begin{acknowledgments}
 We are grateful to Kirill Petrovnin and Shruti Dogra for assistance with the measurements and to Henrik Lievonen for help with data analysis. We  acknowledge financial support from the RADDESS programme (project 328193) of the Academy of Finland and from Grant No. FQXi-IAF19-06 (“Exploring the fundamental limits set by thermodynamics in the quantum regime”) of the Foundational Questions Institute Fund (FQXi), a donor advised fund of the Silicon Valley Community Foundation. This work is part of the Finnish Center of Excellence in Quantum Technology QTF (projects 312296, 336810)
of the Academy of Finland. One of the samples used in this work was produced using the material and technical resources of the Common Use Center 
of the Research and Education Center ``Functional Micro/Nanosystems'' of the Bauman Moscow State 
Technical University.  This work used the experimental facilities of the Low Temperature Laboratory of OtaNano and is part of the European Microkelvin Platform project, EMP (grant agreement no. 824109).
 \end{acknowledgments}

\section*{SUPPLEMENTARY MATERIAL}
 
In this supplementary material we present more details about the experiments, including examples of the measured responses used for the analysis in the main text. Also we validate the proposed protocol by a numerical simulation. Finally, the errors of the protocol are discussed.

\maketitle

\renewcommand{\theequation}{S\arabic{equation}}
\renewcommand{\thefigure}{S\arabic{figure}}

\section{Measurement setup}\label{Setup}

\begin{figure*}[ht]
\includegraphics[width=0.95\textwidth]{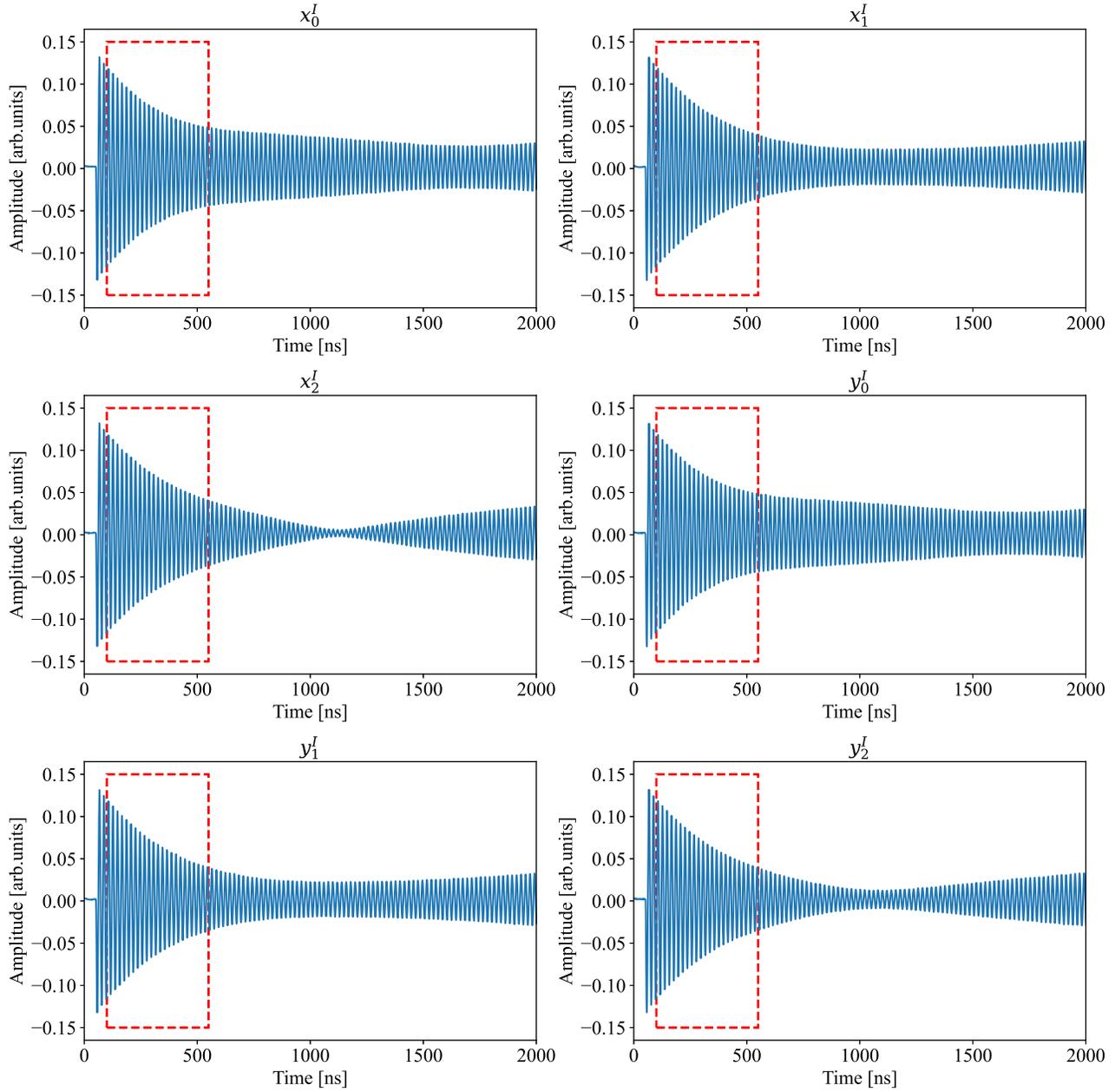}
\caption{\label{responses} Typical measured $I$-quadrature responses after applying the  swapping sequences, listed in Table I of the main text. \textcolor{black}{The measurement pulse has a 2 $\mu$s duration} and we measure in the heterodyne configuration with intermediate frequency (IF) of 50 MHz. The red dashed-line rectangles show the range where the difference between outputs is maximal and used for the analysis.}
\end{figure*}

\textcolor{black}{The readout method in our experiments is based on heterodyne detection, where the IF frequency is ~$50~\mathrm{MHz}$. This has the advantage that it suppresses the power sent to the resonator while it is not probed, which would otherwise be dominated by the LO tone. The digitized signals will therefore oscillate at the IF frequency:  typical responses measured this way are shown in Fig. \ref{responses}. With  single-shot readout the natural thing to do would be to demodulate the pulses and to aggregate the information into a single complex number, then the state readout is equivalent to classifying the measured response into one of $n$ categories, where $n$ is the number of energy levels. However, in an averaged readout scheme one must rely on fitting to extract the populations: a measured response $\varphi$ is decomposed in the basis of pure state responses $\varphi_{g,e,f}$ (obtained by applying no drive pulse, $\pi_{ge}$ and $\pi_{ef}\pi_{ge}$ respectively), as $\varphi = \varphi_g p_g + \varphi_e p_e  +\varphi_f p_f $, and finally the populations $p_{g,e,f}$ can be found by linear regression. In our case the responses $\varphi$ are complex $I + iQ$ signals, and demodulating them does not alter the outcome of the fitting procedure. The red rectangles in figure \ref{responses} show the data range used for the analysis, the starting point is defined by the so-called ring-up time of the readout resonator (in our case it is $t_m>Q_{\rm loaded}/(4f_{r})\approx 100$ ns), while the final point is  restricted by the transmon's lifetime. As the probe pulse duration is comparable with the relaxation times of our transmon, the probe signal correlation increases as a function of time. Therefore in the analysis we imposed a cut-off, chosen to maximize the differences between the calibration responses. The cutoff is set at $450$ ns, which is sufficiently shorter than the measured lifetimes. By using the finite duration of the readout pulse for the linear regression we increase the fidelity of the readout procedure: the inherent time-dependence of the traces enables us to construct an over-determined system of equations for $p_{g,e,f}$ (i.e. each digitizer sample is one pair of equations).\\
In the context of this work, this readout scheme has an important property: the temperature measurement protocol relies on cancelling one of the pure state responses from the averaged one, and the difference lies along the $\varphi_{ij} = \varphi_i - \varphi_j$ ($i,j \in \{g,e,f\}$). If the contribution of the third response is not canceled it would show up as an ellipse instead of a linear dependence in Fig. 1 of the main text (as well as Figs.  \ref{SimulA} and \ref{fig2}). Thus, the oscillating nature of the readout pulses can be used to easily verify that the protocol is implemented properly.}

Each point in Fig. 1 in the main text corresponds to one sample of the $x_0^I(t),x_1^I(t),y_0^I(t),y_0^I(t)$ signals, where $x_0^I(t)-x_1^I(t)$ is shown on the $y$ and $y_0^I(t)-y_0^I(t)$ on the $x$ axis.

 
\textcolor{black}{As in many experiments in circuit QED, thermalization is not perfect, and several mechanisms have been proposed to explain the additional heating -- most notably non-thermal excitations coming from the hotter stages of the dilution refrigerator, infrared radiation, and rf fields coupling into the bias lines and the sample due to imperfect shielding. This can be seen already in Experiment 1, where the effective temperature is 50 mK. In principle, this temperature can be further reduced by a few tens of mK by the use of infrared filters, an additional shield attached to the mixing chamber, and microwave-absorbing coatings. In Experiment 2 the goal is to increase this temperature, which was done by removing 15 dB of attenuation from the mixing chamber. Thus, the radiation coming from the upper stages, either thermal or non-equilibrium,  delivers a higher power to the transmon, and, furthermore, the inner (signal) conductor of this line is now imperfectly thermalized.}

\section{Simulation of the system}\label{Simul_app}
We have done numerical simulations using the Qutip package \cite{Johansson2013} based on the following transmon Hamiltonian \cite{Koch2007}:
\begin{align}\label{Hamiltonian}
\hat{H}=4E_c\left(\hat{n}-n_g\right)^2-E_J\cos \hat{\varphi} + \hbar \omega_r{\hat{a}}^\dag \hat{a}+2\beta e V_{\rm rms}^{0}\hat{n}\left({\hat{a}}^\dag+\hat{a}\right), \nonumber
\end{align}
where $E_c,E_J$ are the Coulomb and Josephson energy of the transmon with the charge number operator $\hat{n}$ and phase operator $\hat{\varphi}$, and $\omega_{r}= 1/\sqrt{L_{r}C_{r}}$ is the resonator's angular frequency, with bosonic operators $\hat{a}^\dag, \hat{a}$. The fourth term describes the fact that the transmon is coupled via a coupling factor $\beta = C_{g}/C_{\Sigma}$ to the vacuum fluctuations of the resonator $V_{\rm rms}^{0}= \sqrt{\hbar \omega_{r}/C_{r}}$.  To this Hamiltonian we add the interaction of a classical drive of angular frequency $\omega_{d}$ and envelope $A_{d}(t)$ with the transmon, and that of a classical probe field of angular frequency $\omega_{p}$ and envelope $A_{p}(t)$ with the resonator

\begin{equation*}
H_{\rm sig}=A_d (t) e^{i\omega_{d}t}\hat{n} + A_p\left(t\right)e^{i\omega_{p}t}\hat{a} + h.c.
\end{equation*}
In our experiments $A_d\left(t\right)$ has a Gaussian envelope and $A_p\left(t\right)$ has a rectangular envelope. To find the initial state before implementing the sequences in Table I of the main text we find the steady-state solution describing the initial thermalized mixed state. 
To do the simulation numerically we should restrict the Hilbert space of the system. We take the four  first states of the transmon qubit ($\left| g \right\rangle,\left| e \right\rangle,\left| f \right\rangle,\left| d \right\rangle $) and the resonator states up to $6$ photons. 
The simulation of our system is based on the numerical solution for the density matrix evolution equation, where dissipation effects are considered in the Lindblad master equation
\begin{equation}\label{dens_eq}
    \dot{\rho} = -\frac{i}{\hbar}[H(t),\rho(t)] + \sum_{k,l = g,e,f}^{k\neq l}  D[L_{kl}]\rho  
    + \sum_{m = \uparrow, \downarrow} D[L_{r}^{m}]\rho , 
\end{equation}
The Lindblad superoperators are defined as $D[L]\rho = (2L\rho L^{\dagger} - L^{\dagger}L\rho - \rho L^{\dagger}L)/2$, where $L$ is a jump operator. For the transmon, we can define raising and lowering operators corresponding to each transition by $\sigma_{kl} =|k\rangle \langle l|$ ~\cite{Kumar2016}. 
Since our protocol involves only population transfers, the pure dephasing rates do not play any role, as they would affect the off-diagonal elements. The Lindblad operators have the following form:
 \begin{equation}
\begin{array}{l}
{L_{eg}} = \sqrt {{\Gamma_{eg}}n_{eg}^{\rm th}}\sigma_{eg}\otimes \mathbb{I}_{r},\\
{L_{fe}} = \sqrt {{\Gamma_{fe}}n_{fe}^{\rm th}} \sigma_{fe}\otimes \mathbb{I}_{r},\\
L_r^{\uparrow} = \sqrt {\kappa n_{r}^{\rm th}} \mathbb{I}_{3}\otimes {a^\dag },\\
{L_{ge}} = \sqrt {{\Gamma_{eg}}\left( {n_{eg}^{\rm th} + 1} \right)} \sigma_{ge}\otimes \mathbb{I}_{r},\\
{L_{ef}} = \sqrt {{\Gamma_{fe}}\left( {n_{fe}^{\rm th} + 1} \right)} \sigma_{ef}\otimes \mathbb{I}_{r},\\
L_r^{\downarrow} = \sqrt {\kappa \left( {n_{r}^{\rm th} + 1} \right)} \mathbb{I}_{3} \otimes a,\\

\end{array}
 \end{equation}
where $\mathbb{I}_{3}$ is the identity 3x3 matrix, $\mathbb{I}_{r}$ is the identity matrix in the Hilbert space of the resonator,
and the number of thermal photons is defined in a standard way:

\begin{eqnarray}
n^{\rm th}_{ij} &=& \frac{1}{\exp{(hf_{ij}/k_{\rm B}T})-1}, \\
n^{\rm th}_{r} &=& \frac{1}{\exp{(hf_{r}/k_{\rm B}T})-1},
\end{eqnarray}
where $i,j\in \{g,e,f\}$. The decay rates $\Gamma_{eg}$ and $\Gamma_{fe}$
are a consequence of the coupling of the transmon with the environment, and the direct decay $f\rightarrow g$ is forbidden by selection rules, therefore $L_{gf}=L_{fg}=0$. The values of 
$\Gamma_{eg}$ and $\Gamma_{fe}$ are found from relaxation measurements, while $\kappa$ is extracted from the loaded quality factor of the resonator, $Q_{\rm load} = 2\pi f_{r}/\kappa$ (obtained from the cavity spectrum with the qubit far off-resonant).
The Lindblad operators $L_{eg}$
and $L_{fe}$ define thermal excitations, while $L_{ge}$ and $L_{ef}$ define induced relaxation into the environment.
Note that the principle of detailed balance for the relaxation rates is satisfied since $n^{\rm th}_{ij}+1 = n^{\rm th}_{ij}\exp{(hf_{ij}/k_{\rm B}T)}$.

\begin{figure}[!ht]
\includegraphics[width=0.95\linewidth]{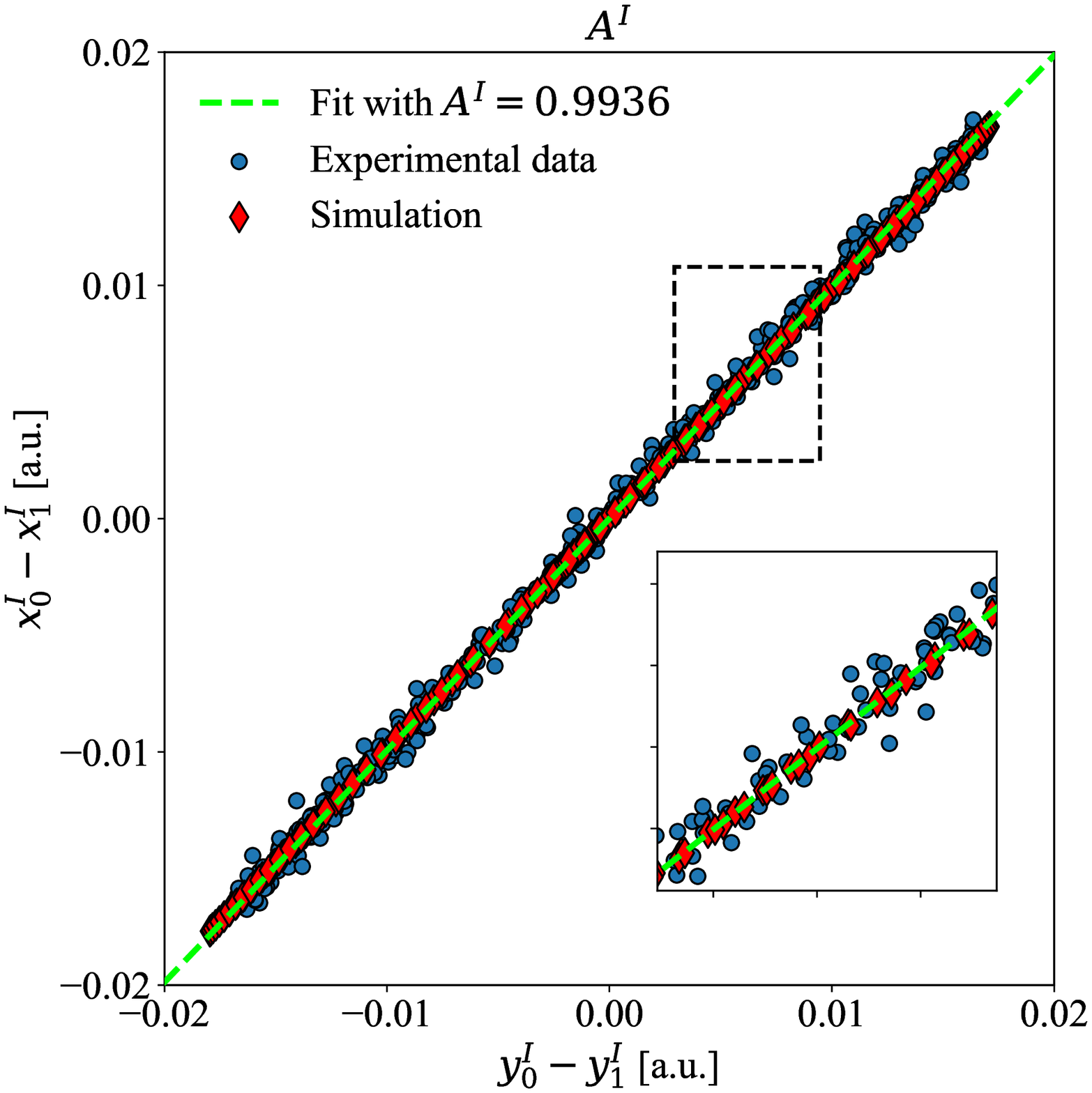}
\caption{\label{SimulA} Differences $x_0^{I}-x_1^{I}$ versus $y_0^{I}-y_1^{I}$
in the measured resonator response measured experimentally (blue dots) compared with those obtained from the simulation (red diamonds) \textcolor{black}{and fitting (green dashed line).} \textcolor{black}{The inset shows a detail of these results.}} 
\end{figure}

The steady-state solution of Eq. (\ref{dens_eq}) gives us thermalized state of the transmon, which is chosen as initial state for the evolution during corresponding sequences. 
To get the proper parameters of $\pi_{01}$ and $\pi_{12}$ pulses we emulated the Rabi-oscillation experiment, while tuning the amplitude, frequency and duration of corresponding Gaussian pulses. 
Once this is done we define the sequences needed for the thermometry and set the readout square pulse at the resonator's frequency for each sequence. 

We can therefore calculate expectation values $\left\langle {a} \right\rangle$ and use real and imaginary part as the quadratures of the measured field. 
The final result is shown in Fig. \ref{SimulA}, where we compare the simulations with the measured values. We obtain a very good agreement between the simulation and  the experiment, as expected, where the latter shows a slightly higher dispersion around the linear fit.

\section{Error estimation}\label{Error}

\begin{figure}[!ht]
\includegraphics[width=0.95\linewidth]{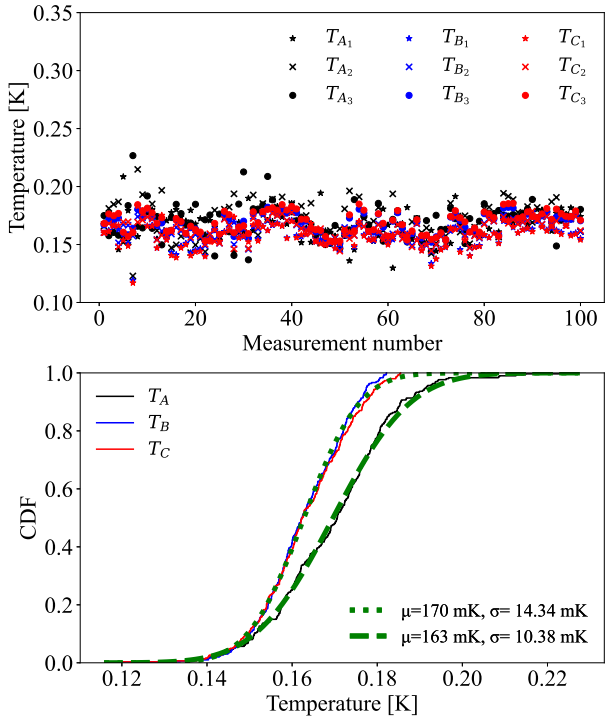}
\caption{\label{fig1} Upper panel: the measured temperatures $T_{A,B,C}$ for all three response vectors as a function of time. Lower panel: plot of the \textcolor{black}{cumulative probability function (CDF)  of the data above.}} 
\end{figure}

In order to estimate the experimental errors of our protocol the following experiment was performed: the temperature was measured 100 times, over the course of $\approx15\mathrm{min}$. We used for this a different sample --  but nominally \textcolor{black}{of the same design and in the same configuration as in Experiment 2} of the main text, with $f_{ge}=6.74$ GHz and $f_{gf}=13.14$ GHz. The results are shown in Fig. \ref{fig1}. 

 As stated in the main text $A$, $B$ and $C$ can be determined in 3 different ways, by taking the prescribed difference, resulting in a signal colinear with one of the following directions: ${\varphi}_g^{I(Q)}-{\varphi}_e^{I(Q)}$, ${\varphi}_g^{I(Q)}-{\varphi}_f^{I(Q)}$ or ${\varphi}_e^{I(Q)}-{\varphi}_f^{I(Q)}$. Unlike in the main text, here the slopes $A$, $B$ and $C$ are determined by fitting the data from both quadratures simultaneously, with no discernible difference. Since all three combinations, shown as different symbols in Fig. \ref{fig1}, coincide this confirms that the pure state responses ${\varphi}_{g,e,f}^{I(Q)}$ were sufficiently distinguishable in our experiment. Even if this were not the case, having two easily distinguishable states is sufficient for estimating the temperature.
 
\begin{figure}[!ht]
\includegraphics[width=0.95\linewidth]{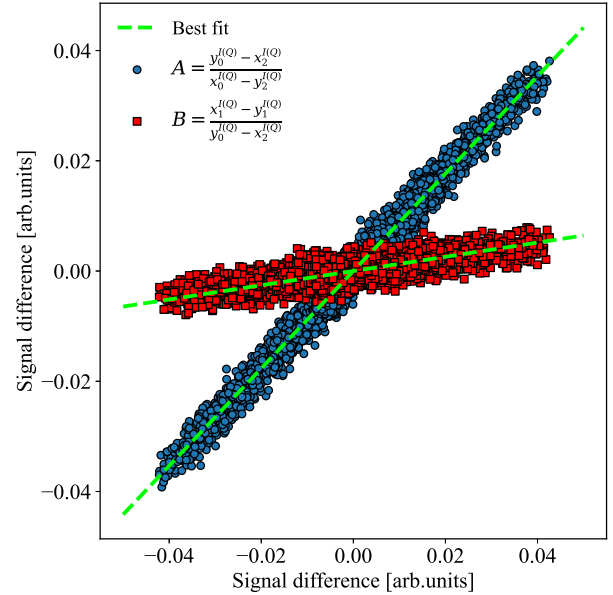}
\caption{\label{fig2} An example of the data used to determine the $A$ and $B$ temperatures, proportional to the ${\varphi}_g^{I(Q)}-{\varphi}_f^{I(Q)}$ response vector.} 
\end{figure}

On the other hand, the values obtained from $A$ differ and have a larger spread. As no clear trend is observed, implying that the temperature was stable during the experiment, we analyzed the statistics of these measurements. The lower panel of \ref{fig1} shows the cumulative probability for $T_A$, $T_B$ and $T_C$, averaged over all three combinations. $T_A$ has a mean of $169.9~\mathrm{mK}$ and a standard deviation of $14.4~\mathrm{mK}$, while $T_B$ and $T_C$ have the same mean of $162.8~\mathrm{mK}$ and a standard deviation of $10.4~\mathrm{mK}$.

\begin{figure*}[!ht]
	\makebox[\textwidth][c]{\includegraphics[width=0.95\linewidth]{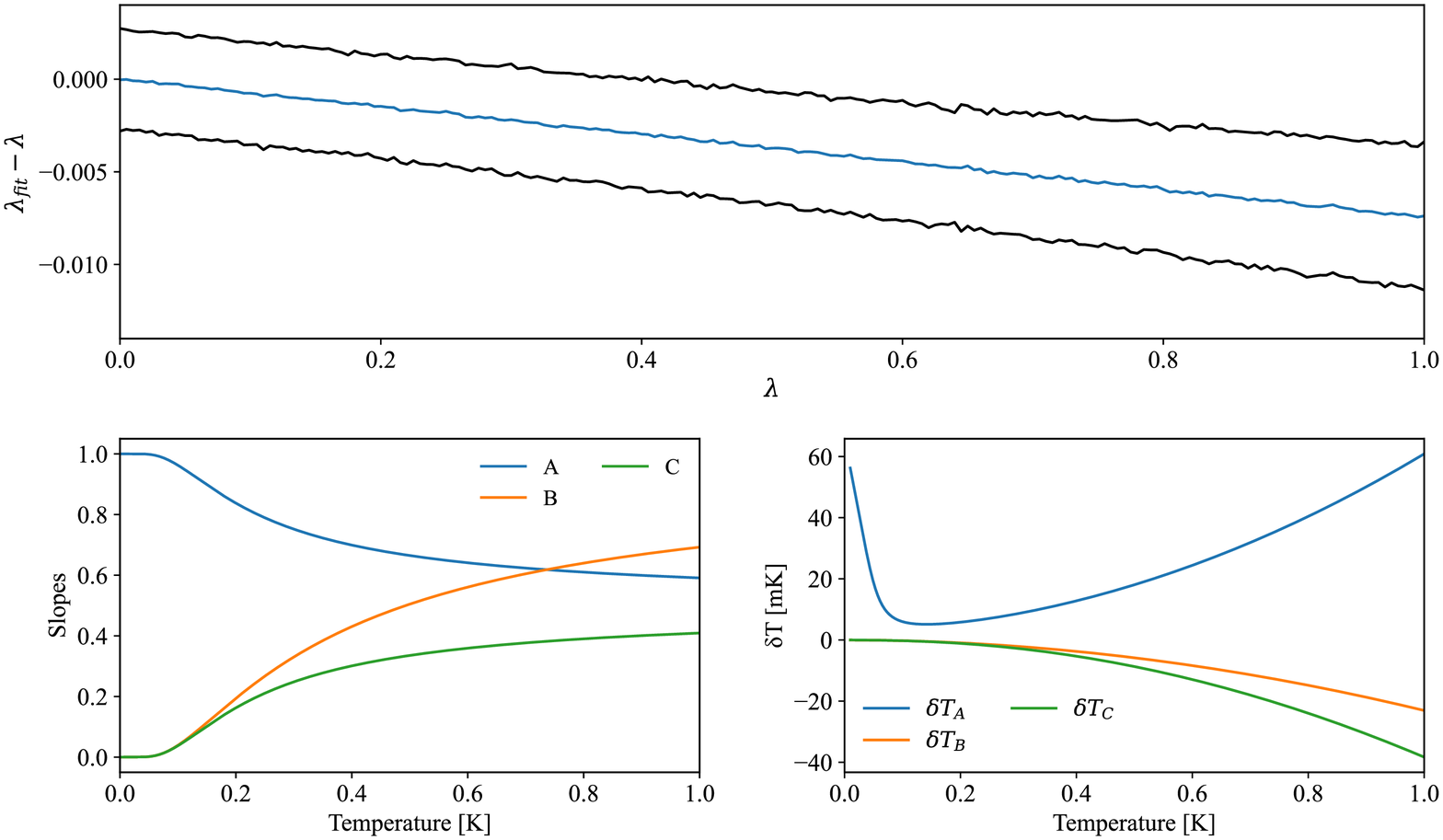}}
	\caption{\label{fig3} Top: the error of the fitted slope $\lambda_{fit}$ versus the actual one $\lambda$, averaged over 1000 numerical experiments. The black lines show the 95\% confidence intervals. Bottom left: the $A$, $B$ and $C$ coefficients as a function of temperature for the transmon frequencies as given at the top of this section. Bottom right: the discrepancy of temperatures $T_{A,B,C}$ caused by the erroneously determined slope.} 
\end{figure*}

We have identified the finite signal to noise ratio of the experimental data as the source of this discrepancy. A typical example of the I/Q data from this data set is shown in Fig. \ref{fig2}: the x and y variables are linearly correlated, the x span is approximately $\pm0.042$, and the noise in both directions is assumed to be Gaussian with an estimated standard deviation of $\approx 0.002$. The best fit slopes are found to be $\lambda_B = 0.1320$ with a 95\% confidence interval of $\{0.129,0.135\}$, and $\lambda_A=0.8834$ with a 95\% confidence interval of $\{0.8799,0.8869\}$. If we model the data as described above, we find that there is a systematic discrepancy between the actual slope and the fitted one, and that it grows linearly with the value of the slope, as shown on the left panel of Fig. \ref{fig3}. It is important to note that for $\lambda\approx 0.88$ the 95\% confidence interval is always below the true value. Together with the temperature dependence of $A$, $B$ and $C$ slopes (low-left panel of \ref{fig3}), this leads to an overestimation of the temperature as measured by $A$ by  $\approx 5\mathrm{mK}$ and a slight underestimation ($\approx 1\mathrm{mK}$) for $B$ and $C$ (right panel of \ref{fig3}).

We conclude that the relatively low signal to noise ratio of the IQ response is predominantly to blame for these discrepancies and for the noise in the temperature measurement. Reducing the readout noise by a factor of two can suppress these errors below $1\mathrm{mK}$.


\section*{References}
\begin{thebibliography}{40}

\bibitem{Rigetti2012}
C. Rigetti, J. M. Gambetta, S. Poletto, B. L. T. Plourde, J. M. Chow, A. D. Córcoles, J. A. Smolin, S. T. Merkel, J. R. Rozen, G. A. Keefe, M. B. Rothwell, M. B. Ketchen, and M. Steffen, Superconducting qubit in a waveguide cavity with a coherence time approaching 0.1 ms, Phys. Rev. B {\bf 86}, 100506(R). 

\bibitem{Oliver2013}
W. D. Oliver and P. B. Welander, Materials in superconducting quantum bits, MRS Bull. {\bf 38}, 816 (2013).

\bibitem{Paik2011}
H. Paik, D. I. Schuster, L. S. Bishop, G. Kirchmair, G. Catelani, A. P. Sears, B. R. Johnson, M. J. Reagor, L. Funzio, L. I. Glazman, S. M. Girvin, M. H. Devoret, and
R. J. Schoelkopf, Observation of high coherence in Josephson junction qubits measured in a three-dimensional circuit qed architecture., Phys. Rev. Lett. {\bf 107}, 240501 (2011).

\bibitem{Geerlings2013}
K. Geerlings, Z. Leghtas, I. M. Pop, S. Shankar, L. Frunzio, R. J. Schoelkopf, M. Mirrahimi, and M. H. Devoret, Demonstrating a driven reset protocol for a superconducting qubit., Phys. Rev. Lett. {\bf 110}, 120501 (2013).

\bibitem{Valenzuela2006}
 S. O. Valenzuela, W. D. Oliver, D. M. Berns, K. K. Berggren, L. S. Levitov, and T. P. Orlando, Microwave-induced cooling of a superconducting qubit, Science {\bf 314}, 1589 (2006).

\bibitem{Tuorila2017}
J. Tuorila, M. Partanen, T. Ala-Nissila and M. Möttönen, Efficient protocol for qubit initialization with a tunable environment, npj Quantum Information {\bf 3}, 27 (2017).

\bibitem{Riste2012}
D. Ristè, J. G. van Leeuwen, H.-S. Ku, K. W. Lehnert, and L. DiCarlo, Initialization by Measurement of a Superconducting Quantum Bit Circuit, Phys. Rev. Lett. {\bf 109}, 050507 (2012).

\bibitem{Egger2018}
D.J. Egger, M. Werninghaus, M. Ganzhorn, G. Salis, A. Fuhrer, P. Müller, and S. Filipp, Pulsed Reset Protocol for Fixed-Frequency Superconducting Qubits, Phys. Rev. Applied {\bf 10}, 044030 (2018).

\bibitem{Magnard2018}
P. Magnard, P. Kurpiers, B. Royer, T. Walter, J.-C. Besse, S. Gasparinetti, M. Pechal, J. Heinsoo, S. Storz, A. Blais, and A. Wallraff, Fast and Unconditional All-Microwave Reset of a Superconducting Qubit, Phys. Rev. Lett. {\bf 121}, 060502 (2018).

\bibitem{Campagne2020}
P. Campagne-Ibarcq, A. Eickbusch, S. Touzard, E. Zalys-Geller, N. E. Frattini, V. V. Sivak, P. Reinhold, S. Puri, S. Shankar, R. J. Schoelkopf, L. Frunzio, M. Mirrahimi and  M. H. Devoret,Quantum error correction of a qubit encoded in grid states of an oscillator, Nature {\bf 584}, 368 (2020).

\bibitem{Reed2012}
M. D. Reed, L. DiCarlo, S. E. Nigg, L. Sun, L. Frunzio, S. M. Girvin, and R. J. Schoelkopf, Realization of three qubit quantum error correction with superconducting circuits., Nature (London) {\bf 482}, 382 (2012).

\bibitem{Vijay2011}
R. Vijay, D. H. Slichter, and I. Siddiqi, Observation of quantum jumps in a superconducting artificial atom, Phys. Rev. Lett. {\bf 106}, 110502 (2011).

\bibitem{deLange2014}
G. de Lange, D. Ristè, M. Tiggelman, C. Eichler, L. Tornberg, G. Johansson, A. Wallraff, R. Schouten, and L. DiCarlo, Reversing quantum trajectories with analog feedback, Phys. Rev. Lett. {\bf 112}, 080501 (2014).

\bibitem{Lin2013}
Z. R. Lin, K. Inomata, W. D. Oliver, K. Koshino, Y. Nakamura, J. S. Tsai, and T. Yamamoto, Single-shot readout of a superconducting flux qubit with a flux-driven Josephson parametric amplifier, Appl. Phys. Lett. {\bf 103}, 132602 (2013).

\bibitem{Abdo2014}
B. Abdo, K. Sliwa, S. Shankar, M. Hatridge, L. Frunzio, R. Schoelkopf, and M. Devoret, Josephson directional amplifier for quantum measurement of superconducting circuits, Phys. Rev. Lett. {\bf 112}, 167701 (2014).

\bibitem{Corcolesa2011}
A. D. Córcolesa), J. M. Chow, J. M. Gambetta, C. Rigetti, J. R. Rozen, G. A. Keefe, M. B.Rothwell, M. B. Ketchen, and M. Steffen, Protecting superconducting qubits from radiation , Appl. Phys. Lett. {\bf 99}, 181906 (2011).

\bibitem{Wenner2013}
J. Wenner, Yi Yin, Erik Lucero, R. Barends, Yu Chen, B. Chiaro, J. Kelly, M. Lenander, Matteo Mariantoni, A. Megrant, C. Neill, P. J. J. O’Malley, D. Sank, A. Vainsencher, H. Wang, T. C. White, A. N. Cleland, and John M. Martinis, Excitation of Superconducting Qubits from Hot Nonequilibrium Quasiparticles, Phys. Rev. Lett. {\bf 110}, 150502  (2013).

\bibitem{Barends2011}
R. Barends, J. Wenner, M. Lenander1, Y. Chen, R. C. Bialczak, J. Kelly, E. Lucero, P. O’Malley, M. Mariantoni, D. Sank, H. Wang1 T. C. White, Y. Yin, J. Zhao, A. N. Cleland, J. M. Martinis, and J. J. A. Baselmans, Minimizing quasiparticle generation from stray infrared light in superconducting quantum circuits, Appl. Phys. Lett. {\bf 99}, 113507 (2011).

\bibitem{Serniak2018} K. Serniak, M. Hays, G. de Lange, S. Diamond, S. Shankar, L. D. Burkhart, L. Frunzio, M. Houzet, and M. H. Devoret, 
Hot non-equilibrium quasiparticles in transmon qubits, Phys. Rev. Lett. {\bf 121}, 157701 (2018).

\bibitem{Pekola2015} J. P. Pekola, Towards quantum thermodynamics in electronic circuits, Nat. Phys. {\bf 11}, 118 (2015).

\bibitem{Silveri2017} M. P. Silveri, J. A. Tuorila, E. V. Thuneberg, and G. S. Paraoanu, Quantum systems under frequency modulation, Rep. Progr. Phys. {\bf 80}, 056002 (2017).

\bibitem{Marco2021} M. Cattaneo and G. S. Paraoanu,  
Engineering dissipation with resistive elements in circuit quantum electrodynamics, arXiv: 2103.16946, to appear in Advanced Quantum Technologies. 

\bibitem{Kulikov2020}
A. Kulikov, R. Navarathna, and A. Fedorov Measuring Effective Temperatures of Qubits Using Correlations, Phys. Rev. Lett. {\bf 124}, 240501  (2020).

\bibitem{Johnson2012}
J. E. Johnson,C. Macklin, D. H. Slichter, R. Vijay, E. B. Weingarten, J. Clarke,and I. Siddiqi, Heralded State Preparation in a Superconducting Qubit, Phys. Rev. Lett. {\bf 109}, 050506 (2012).

\bibitem{Riste2012_2}
D. Ristè, C. C. Bultink, K. W. Lehnert, and L. DiCarlo, Feedback Control of a Solid-State Qubit Using High-Fidelity Projective Measurement, Phys. Rev. Lett. {\bf 109}, 240502 (2012).

\bibitem{Krantz2016}
P.  Krantz, A. Bengtsson, M. Simoen, S. Gustavsson, V. Shumeiko, W. D. Oliver, C. M. Wilson, P. Delsing and J. Bylander, Single-shot read-out of a superconducting qubit using a Josephson parametric oscillator, Nature Communications {\bf 7}, 11417 (2016). 

\bibitem{Jin2015}
X. Y. Jin, A. Kamal, A. P. Sears, T. Gudmundsen, D. Hover, J. Miloshi, R. Slattery, F. Yan, J. Yoder, T. P. Orlando, S. Gustavsson, and W. D. Oliver, Thermal and Residual Excited-State Population in a 3D Transmon Qubit, Phys. Rev. Lett. {\bf 114}, 240501 (2015).

\bibitem{Simone2020} M. Scigliuzzo, A. Bengtsson, J.-C. Besse, A. Wallraff, P. Delsing, and S. Gasparinetti, Primary Thermometry of Propagating Microwaves in the Quantum Regime, Phys. Rev. X {\bf 10}, 041054 (2020).

\bibitem{Biancetti2010}
R. Bianchetti, S. Filipp, M. Baur, J. M. Fink, C. Lang, L. Steffen, M. Boissonneault, A. Blais, and A. Wallraff, Control and Tomography of a Three Level Superconducting Artificial Atom, Phys. Rev. Lett. {\bf 105}, 223601 (2010)

\bibitem{deming1943}
W. E. Deming, Statistical adjustment of data, Wiley (1943).

 \bibitem{Clerk2010}
A. A. Clerk, M. H. Devoret, S. M. Girvin, F. Marquardt, and R. J. Schoelkopf, Introduction to quantum noise, measurement, and amplification, Rev. Mod. Phys. {\bf 82}, 1155 (2010).

\bibitem{Karimi2016} B. Karimi and J. P. Pekola, Otto refrigerator based on a superconducting qubit: Classical and quantum performance, Phys. Rev. B {\bf 94}, 184503 (2016).

\bibitem{Sina2021} S.H. Raja, S. Maniscalco, G.S. Paraoanu, J.P. Pekola, and N.L. Gullo, Finite-time quantum Stirling heat engine, New J. Phys. {\bf 23}, 033034 (2021).

\bibitem{Krinner2019}
S. Krinner, S. Storz, P. Kurpiers, P. Magnard, J. Heinsoo, R. Keller, J. Lutolf, C. Eichler and A. Wallraff
Engineering cryogenic setups for 100-qubit scale superconducting circuit systems, EPJ Quantum technology {\bf 6}, 2 (2019).

\bibitem{Grunhaupt2018} L. Grünhaupt, N. Maleeva, S. T. Skacel, M. Calvo, F. Levy-Bertrand, A. V. Ustinov, H. Rotzinger, A. Monfardini, G. Catelani, and I. M. Pop, Loss Mechanisms and Quasiparticle Dynamics in Superconducting Microwave Resonators Made of Thin-Film Granular Aluminum, Phys. Rev. Lett. {\bf 121}, 117001.

\bibitem{Riste2013}
D. Ristè, C. C. Bultink, M. J. Tiggelman, R. N. Schouten, K. W. Lehnert and L. DiCarlo,
Millisecond charge-parity fluctuations and induced decoherence in a superconducting transmon qubit, Nat. Comm. {\bf 4}, 1913 (2013).

\bibitem{Weides2019} S. Schl\"or, J. Lisenfeld, C. M\"uller, A. Bilmes, A. Schneider, D. P. Pappas, A. V. Ustinov, and M. Weides, Correlating Decoherence in Transmon Qubits: Low Frequency Noise by Single Fluctuators,
Phys. Rev. Lett. {\bf 123}, 190502 (2019).

\bibitem{Bylander2019} J. J. Burnett, A. Bengtsson, M. Scigliuzzo, D. Niepce, M. Kudra, P. Delsing, and  J. Bylander, Decoherence benchmarking of superconducting qubits, npj Quantum Information {\bf 5}, 54 (2019)

\bibitem{Combes2021} C. R. H. McRae, G. M. Stiehl,  H. Wang, S.-X. Lin, S. A. Caldwell, D. P. Pappas, J. Mutus, and J. Combes, Reproducible coherence characterization of superconducting quantum devices,  Appl. Phys. Lett. {\bf 119}, 100501 (2021).

\bibitem{Catelani2011} G. Catelani, J. Koch, L. Frunzio, R. J. Schoelkopf, M. H. Devoret,and L. I. Glazman, Quasiparticle Relaxation of Superconducting Qubits in the Presence of Flux, Phys. Rev. Let. 
{\bf 106}, 077002 (2011).

\bibitem{Martinis2009} J. M. Martinis, M. Ansmannand J. Aumentado, Energy Decay in Superconducting Josephson-Junction Qubits from Nonequilibrium Quasiparticle Excitations, Phys. Rev. Lett. {\bf 103}, 097002 (2009).


\bibitem{Johansson2013}
J. R. Johansson, P. D. Nation, and F. Nori, QuTiP 2: A Python framework for the dynamics of open quantum systems, Comp. Phys. Comm. {\bf 184}, 1234 (2013). 

\bibitem{Koch2007}
J. Koch, T. M. Yu, J. Gambetta, A. A. Houck, D. I. Schuster, J. Majer, A. Blais, M. H. Devoret, S. M. Girvin and R. J. Schoelkopf, Charge-insensitive qubit design derived from Cooper pair box, Phys. Rev. A {\bf 76}, 042319 (2007)

\bibitem{Kumar2016} K.S. Kumar, A. Vepsäläinen, S. Danilin, and G. S.  Paraoanu, Stimulated Raman adiabatic passage in a three-level superconducting circuit,
Nat. Commun. {\bf 7}, 1 (2016).



\end{thebibliography}
\end{document}